\documentclass[12pt, preprint]{aastex}
\newcommand\ea{et al.\ }

\def\chandra{{\it Chandra}}
\newcommand\psc{\ifmmode{\rm\,cm^{-2}}\else{${\rm\,cm^{-2}}$}\fi}
\shorttitle{Extreme Iron Lines in AGN}
\shortauthors{Levenson et al.}
\slugcomment{Accepted for publication in The Astrophysical Journal Letters}

\begin{document}
\title{Extreme X-ray Iron Lines in Active Galactic Nuclei}

\author{N. A. Levenson\altaffilmark{1,2}, 
J. H. Krolik\altaffilmark{2}, 
P. T. \.Zycki\altaffilmark{3},
T. M. Heckman\altaffilmark{2}, 
K. A. Weaver\altaffilmark{2,4}, 
H. Awaki\altaffilmark{5},
and
Y. Terashima\altaffilmark{6,7}}

\altaffiltext{1}{Department of Physics and Astronomy, University of Kentucky, Lexington, KY 40506; levenson@pa.uky.edu}
\altaffiltext{2}{Department of Physics and Astronomy, Bloomberg Center, Johns Hopkins University, Baltimore, MD 21218}
\altaffiltext{3}{Nicolaus Copernicus Astronomical Center, Bartycka 18, 00-716 Warsaw, Poland}
\altaffiltext{4}{Code 662, NASA/GSFC, Greenbelt, MD 20771}
\altaffiltext{5}{Department of Physics, Faculty of Science, Ehime University,
Bunkyo-cho, Matsuyama, Ehime 790-8577, Japan}
\altaffiltext{6}{Institute of Space and Astronautical Science, 3-1-1 Yoshinodai, 
Sagamihara, Kanagawa 229-8510, Japan}
\altaffiltext{7}{Department of Astronomy, University of Maryland, College Park, MD 20742}

\begin{abstract}
We analyze X-ray spectra of heavily obscured ($N_H > 10^{24} \psc$)
active galaxies obtained with \chandra{}, concentrating on the 
iron K$\alpha$ fluorescence line.
We measure very large equivalent widths in most cases,
up to 5 keV in the most extreme example. 
The geometry of an obscuring torus of material near the 
active galactic nucleus (AGN) determines the Fe emission,
which we model as a function of
torus opening angle, viewing angle, and optical depth.  
The starburst/AGN composite galaxies in this sample 
require small opening angles.
Starburst/AGN composite galaxies in general 
therefore present few direct lines of sight to their central
engines.  
These composite galaxies are common, and their large covering fractions
and heavy obscuration effectively hide their
intrinsically bright X-ray continua.
While few distant obscured AGNs have been identified,
we propose to exploit their signature large Fe K$\alpha$ equivalent
widths to find more examples in X-ray surveys.
\end{abstract}

\section{Iron Lines in the X-ray Spectra of AGN}

The strongest line in the X-ray spectrum of an active
galactic nucleus (AGN) at moderate energies 
($4\lesssim E \lesssim 10$ keV) is due to
iron fluorescence, particularly 
``neutral'' Fe K$\alpha$  at 6.4 keV,
from Fe less ionized than \ion{Fe}{17}.
If both the continuum and the Fe-emitting region are viewed directly
(the ``Type 1" view), the equivalent width (EW) 
is small, typically less than 200 eV.
As \cite{Kro87} pointed out, the EW can increase greatly if the
fluorescing material is exposed to a stronger continuum than
the observer detects.  In this case (the ``Type 2" view),
an obscuring ``torus'' of material near the active
nucleus blocks direct views of the central 
engine along the line of sight.
X-ray observations of Seyfert galaxies generally
support this unification scenario.
Seyfert 1s typically exhibit EW $\approx 150$ eV \citep{Nan94}, while the
EWs in Seyfert 2s reported hitherto tend to be larger and distributed 
more broadly, ranging from about 100 eV to 1 keV \citep{Tur97}.

Seen from the ``Type 2" view, the K$\alpha$ EW
depends strongly on the torus geometry and total column density.
Previous theoretical calculations \citep*{Awa91,Ghi94,Kro94} 
have concentrated on producing EW $\lesssim 1$ keV,
consistent with earlier observations.  They indicated,
however, that still larger EWs might result from Compton-thick
tori (i.e., $N_H > 10^{24}\psc$ toward the nucleus)
with special geometries.

Motivated by this suggestion and the recent discovery in
\chandra{} X-ray Observatory (\chandra) observations of several
much stronger lines \citep[e.g.,][]{Sam01}, 
in this work, we examine the Fe K$\alpha$ properties of active 
galaxies that have previously been identified as Compton thick.
Our sample, listed in Table \ref{tab:ew}, comprises all such
galaxies \chandra{} has observed for which
published results, archival data, or proprietary data are 
available. 
Most of these are classified as Seyfert 2 galaxies.
NGC 4945 lacks the requisite optical emission line signature of Seyfert
galaxies, but X-rays reveal its active nucleus \citep{Iwa93}.
M51 is exceptional in this group for its low luminosity, 
and the source identified in the \chandra{} Deep Field South, CDF-S 202, 
is exceptional for its
high luminosity, but they both fulfill the broad selection criteria.
Although this sample is not complete and the data are heterogeneous, 
these measurements  
suggest that
extremely large Fe equivalent widths are common in such heavily
obscured AGN.  

\section{Observations and Spectral Fitting\label{sec:fit}}

All of the sample members were observed with
the Advanced CCD Imaging Spectrometer (ACIS),
and the Circinus galaxy and Mrk 3 
were observed with the transmission gratings in place.
See \citet{Wei00}, G. Garmire et al., in preparation,
and \citet{Can00} 
for more information on \chandra, ACIS, and the transmission gratings,
respectively.  

\citet{Sak00} analyze the \chandra{} spectra of 
Mrk 3. 
In order to examine the Fe lines in detail, we obtained these
data from the \chandra{} Data Archive at the \chandra{} X-ray Center
and extracted the  
High-Energy Transmission Grating Spectrometer spectrum 
following the method of \citet{Yaq01}, applying updated calibrations.
Using XSPEC \citep{Arn96}, we fit the spectrum  from 4--8 keV, 
avoiding the complication
of many emission lines at softer energies while providing enough 
high-energy coverage to fit the H-like Fe and neutral Fe K$\beta$
near 7 keV.  We modeled the continuum as a strongly-absorbed
power law, adding an unabsorbed power law constrained to have the
same photon index to represent the scattered contribution.
These data do not strongly constrain the intrinsic column density,
but the EW is relatively insensitive to this uncertainty. 
(These data are in fact consistent with Compton thin obscuration;
$N_H = 6.9 (+2.4, -3.9) \times 10^{23} \psc$.)  
Three Gaussians, which represent the Fe lines, are significant, 
with the neutral K$\alpha$ 
EW$= 0.71$ keV.

M51, NGC 4945, NGC 5135, and NGC 7130,
were all observed with \chandra's
back-illuminated S3 ACIS detector.  
We obtained the data
on NGC 4945  and M51 from the \chandra{} archive. 
See Madejski et al., in preparation, for more complete analysis
of the \chandra{} observation of NGC 4945.  
\citet{Ter01} present results from the observation of M 51,
but we reanalyze the data here 
in order to utilize updated calibrations.
The other two data sets are part of our program on
starburst/AGN composite galaxies. 
We reprocessed all data from original Level 1 event files and
applied the latest gain corrections, from 
2001 July 31. 
We examined the lightcurves and 
excluded times during the NGC 4945
observation where flares were significant.
We detected no
significant flares in the other data.
In each case, we extracted the spectrum from a circular region
of radius about $2\arcsec$ and measured the local background
in nearby source-free regions.
Except in the case of NGC 4945, we did not bin the data, in order to retain
the highest-energy photons.
While the very soft X-ray spectra require thermal emission components,
above 4 keV only the power-law continuum and Fe emission lines
are significant.  A single unabsorbed  continuum component
is required, which we interpret as the reprocessed contribution.
The best-fitting continuum in each of these four spectra is relatively flat,
consistent with pure Compton reflection.

Table \ref{tab:ew} summarizes these results, 
listing the measured central energy and
the equivalent width 
of the neutral 
Fe K$\alpha$ line for each sample member. 
We also record the continuum photon index, $\Gamma$, of the fits.
The intrinsic continuum is strongly reprocessed
in these obscured galaxies.  While the  
simple power laws we apply are reasonable over the restricted 
energy ranges of the model fits, we caution that they are not appropriate
descriptions over large energy ranges (e.g., for $E = 0.5$--$30$ keV).
References in Table \ref{tab:ew} include the
identification of the 
sample members' large column densities and published 
\chandra{} measurements, which we adopt for
Circinus, NGC 1068, CDF-S 202, and IRAS 09104+4109.
In these last two cases, \citet{Nor02} and \citet{Iwa01}
apply reflection models, and the 
tabulated values of $\Gamma$
represent the intrinsic source spectra.
Thus, these two 
are steeper than the directly-measured continua 
of the other highly obscured examples, although their observed
continua are similarly flat.

\section{Fe Line Models\label{sec:model}}

\begin{figure}
\includegraphics[width=6in]{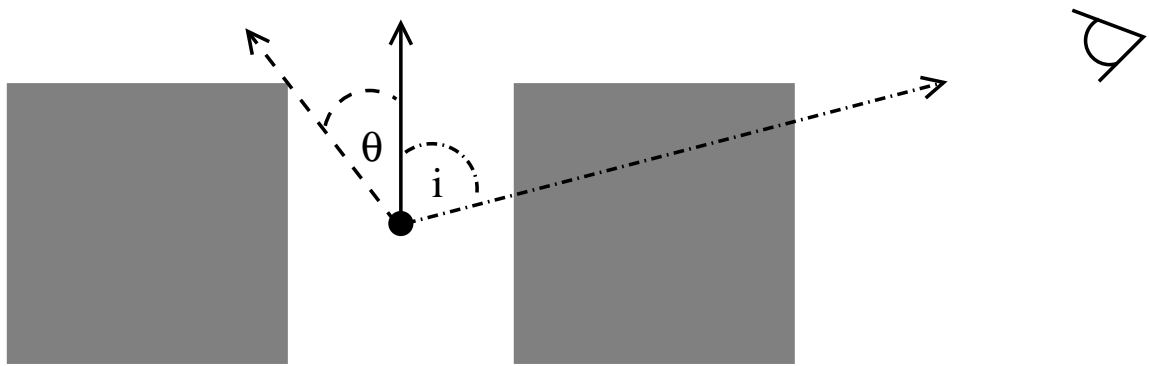} 
\caption{\label{fig:cartoon}
Cartoon of model geometry. The square torus cross-section 
is shaded, and the solid line marks the torus symmetry axis.  
The torus opening angle, $\theta$, and the
viewing angle, $i$, are indicated along dashed and dot-dashed
lines, respectively. 
}
\end{figure}

The Fe K$\alpha$ EWs of the Compton thick AGNs in this sample
tend to be extremely large, as expected when
the intrinsic AGN continuum is fully
hidden from direct view.  
We model the geometric dependence of 
continuum reprocessing with Monte Carlo simulations,
which \citet{Kro94} describe more completely.
Although we introduce no new physics in these calculations, the
current observations demonstrate that 
more extreme configurations are relevant and must be evaluated.
We consider a uniform torus of neutral material having a square cross-section
of Thomson optical depth, $\tau$, in each direction,
and opening angle $\theta$
(Figure \ref{fig:cartoon}).  
We consider all viewing angles, $i$, including Type 1 AGN orientations,
where $i < \theta$.  
We adopt intrinsic $\Gamma = 1.9$ and solar Fe abundance in
the basic model and discuss additional variations below.
We do not consider any scattered or direct contribution from 
ionized material that may be present outside the torus cavity
or near the central source, 
but these simplifications do not significantly affect the EW.

\begin{figure}
\includegraphics[width=4in]{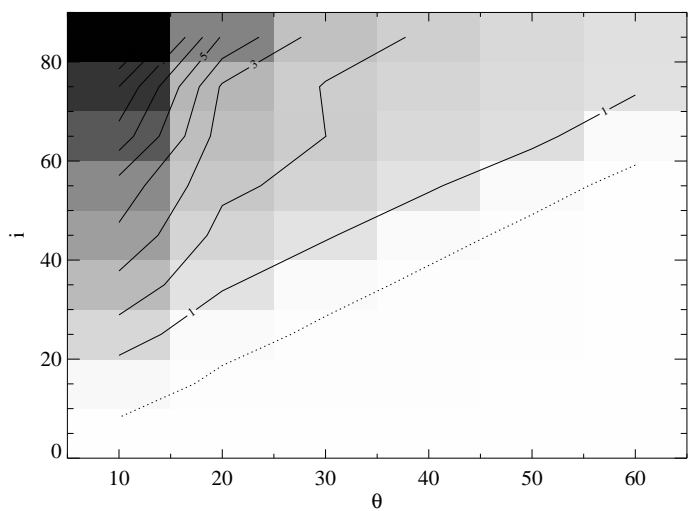} 
\caption{\label{fig:tau4}
EW as a function of $\theta$ and $i$, for $\tau = 4 $ and $\Gamma = 1.9$.
The image is scaled linearly from 0 (white) to 10 keV (black).
The dotted contour is at 100 eV, and the solid contours are scaled
linearly in steps of
1 keV, beginning at 1 keV.
Both $\theta$ and $i$ are indicated in degrees.
}
\end{figure}

The EW rises smoothly with decreasing
$\theta$ and increasing $i$ for constant $\tau$, as
Figure \ref{fig:tau4} illustrates for $\tau = 4$.
For $i \leq \theta$, the EW we calculate falls below $\approx 100$~eV
because we model only the torus contribution; the accretion disk,
visible when $i < \theta$, would add to the EW in that case.

\begin{figure}
\includegraphics[width=4in]{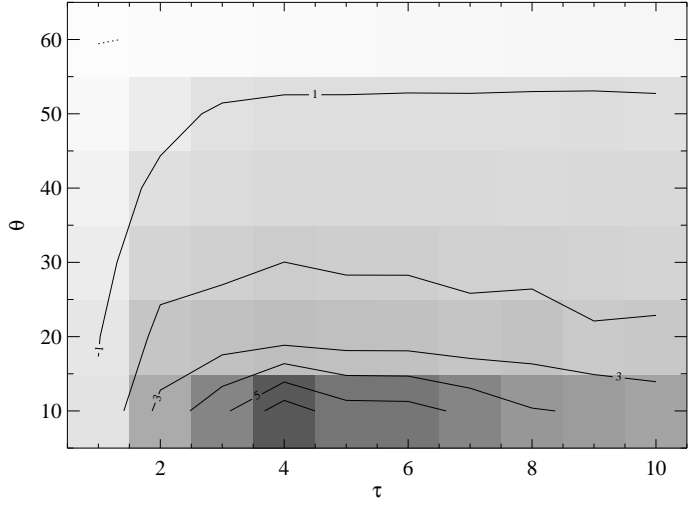} 
\caption{\label{fig:iang65}
EW as a function of $\tau$ and $\theta$, for $i = 65^\circ$.
The image and contour scales are the same as in Fig. \ref{fig:tau4}.
}
\end{figure}

Figure \ref{fig:iang65} shows
the variation of EW in the $\tau-\theta$ plane for $i =  65^\circ$.
For fixed $\tau$, the EW is greatest when $\theta$ is smallest,
but EW does not increase monotonically with $\tau$.
To achieve EW $> 2$ keV, $\tau > 1$ is required; the
greatest EWs are found for $\tau \simeq 4$.

\begin{figure}
\includegraphics[width=4in]{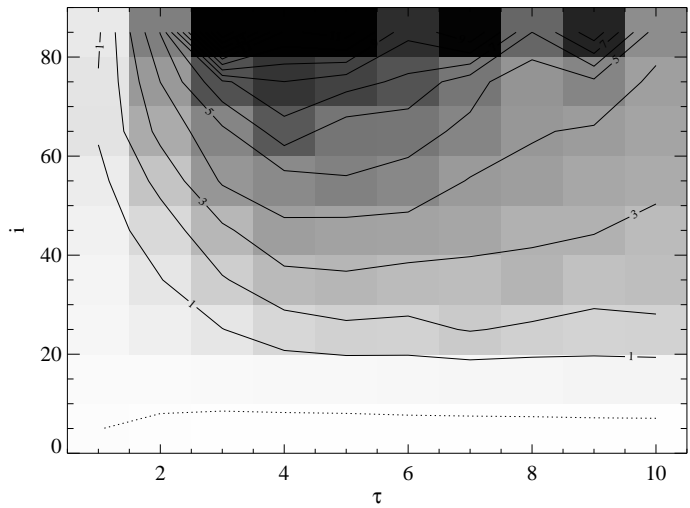} 
\caption{\label{fig:theta10}
EW as a function of $\tau$ and $i$, for $\theta = 10^\circ$.
The image and contour scales are the same as in Fig. \ref{fig:tau4}.
}
\end{figure}

The case of $\theta = 10^\circ$ (Figure \ref{fig:theta10})
demonstrates that extremely large EWs may arise 
over a wide range of viewing angles and optical depths for
sufficiently small torus opening angle.
While EW increases with $i$ for a given $\tau$,
moderate values of $\tau$ and large values of $i$ produce
the largest EWs, which some of these Compton-thick AGNs
require.

These trends may be understood qualitatively as the interaction of several
processes.  Smaller opening angle means the torus obscures a larger solid
angle around the nucleus, hence more Fe K-edge photons are captured and
generate K$\alpha$ photons.  Viewing angles closer to the equatorial plane
see larger equivalent widths in the K$\alpha$ line because the continuum
is most severely attenuated along that direction.  Large, but not too large,
optical depth promotes K$\alpha$ production by absorbing more hard X-rays:
the continuum at K$\alpha$ is suppressed, while photons well above the
Fe K-edge are absorbed and trigger fluorescence.  Compton optical depths
more than $\simeq 4$ diminish the equivalent width because the optical depth
from the points of K$\alpha$ production to the surface is too large for the
fluorescence photons to escape.

Variations in spectral shape, torus cross-section, and Fe abundance
create changes in the EW that are only on the order
of typical measurement errors.
A shallower slope tends to increase the EW,
because relatively more high-energy photons penetrate
deeper into the torus, and the
Fe-fluorescing region is then closer to the observer.
Adopting $\Gamma = 1.7$, for example, the EW changes by
less than 5\% for $i \le 65^\circ$, and up to 30\% at
$i=75^\circ$ for $\theta = 30^\circ$ and $\tau = 2 $.

Previous calculations have
emphasized the effect of Fe abundance in order to 
account for large EWs, but 
abundance variations are only a secondary effect.
With twice-solar abundances, for example, the EW
increases up to 60\% in the edge-on geometry.
In contrast, purely geometric (and physically plausible)
variations  can alter the EW by an order of magnitude.

The Fe line luminosity is correlated with the intrinsic
AGN luminosity.
For a given torus geometry, the K$\alpha$ luminosity is relatively
insensitive to viewing geometry, typically varying 
by factors of 5 over all viewing angles,  with departures to
factors of 10 only in a small minority of cases. 

We constrain the geometries of the
obscuring regions of
the galaxies with the largest EWs with these models.
M51 requires $\theta \le 20^\circ$; for $\theta = 10^\circ$,
a range of optical depth $\tau \ge 2$ and
viewing angle $i \ge 65^\circ$ are consistent. 
NGC 5135 also favors
$\theta \le 20^\circ$,
allowing a range of $i \ge 35^\circ$, depending on $\tau$. 
We note that ionized Fe lines are not detected in these
two galaxies. Their large neutral covering fractions suggest
that even if an ionized region is present,  the neutral
material may completely block it from our view.

\section{Significance\label{sec:signif}}

Larger EWs require more extreme geometries, with
torus opening angle $\theta \lesssim 30^\circ$ for EW$\gtrsim 2$ keV.
In this sample, the galaxies that exhibit the largest EWs have
concentrated circumnuclear starbursts, while in the only galaxy
certainly lacking a starburst, Mrk 3, the K$\alpha$ EW is relatively small.
This may be causal---the mechanical energy of
a starburst, input as stellar winds and supernovae,
may inflate the torus so that it covers a greater solid angle
around the nucleus \citep[e.g., ][]{Wad02}.

The compact starbursts of NGC 5135 and NGC 7130 are evident from both their
vacuum ultraviolet spectra, which show absorption features formed
in the winds and photospheres of massive stars, and from their optical spectra
where the high-order Balmer series and He I lines are observed in absorption
\citep{Gon98}.
The nuclear 
spectrum of M51 reveals high-order Balmer lines in absorption, also
characteristic of a young stellar population \citep{Hec80}.
The 100-pc scale
starburst of NGC 4945 is detected in Pa$\alpha$ images \citep{Mar00},
while a 200-pc ring of H$\alpha$ and Br$\gamma$
is evidence of the young
stars around the nucleus of Circinus 
\citep{Elm98,Mai98}.
The unpolarized nuclear spectrum of IRAS 09104+4109 exhibits a broad
feature around \ion{He}{2} $\lambda 4686$, attributable to Wolf-Rayet stars
\citep{Tra00}.
NGC 1068's starburst is large, and this extended (kpc-scale)
star-formation region \citep{Tel88} therefore does not shape the
immediate environment of the central engine.  Instead, the AGN
significantly affects the conditions of the nuclear region, as the
prominent ionized Fe lines illustrate in this case \citep{Uen94}.

Most recent investigations of the significance of starbursts in active
galaxies have concentrated on Type 2 AGNs
\citep*[e.g.,][]{Gon01} 
for the practical reason
that the starburst signatures are more evident 
when the glare of the central engine is blocked.
Our explanation for the exceptionally large K$\alpha$ EWs seen
in starburst/AGN composites---very small opening angles---implies
that any particular AGN with a starburst is more 
likely to be seen as Type 2,
so few genuine starburst/Type 1 AGNs exist.

Starburst/Type 2 AGN composite galaxies are common.
Approximately half of all Seyfert 2s contain circumnuclear
starbursts \citep{Gon01}.  The composite galaxies are
also preferentially more obscured than their ``pure'' counterparts,
which lack starbursts \citep{LWH01j}.
The starbursts themselves are responsible for some of this obscuration.
With typical star-formation rates per unit area
$SFR \approx 50$--$100 M_\sun {\rm \, kpc^{-2} \, yr^{-1}}$
in the central 100 pc of composite galaxies \citep{Gon98}, 
the correlation of \citet{Ken98} 
implies the mean gas column density in the star-forming
regions is about $10^{24} \psc$.
The causal connection of large column density and star formation 
may then work both ways to enhance EW: 
strong starbursts arise where a large reservoir of material is available,
and the starburst may also inflate the torus.
Thus, in a significant fraction of AGNs,
accretion occurs behind
large obscuring column densities that cover most lines of sight.

In order to reproduce the observed spectrum of the cosmic X-ray background,
synthesis models  \citep{Set89,Com95}  
require a large population of obscured AGNs. 
An outstanding problem with these models has been
the small number of obscured luminous AGNs---Type 2 quasars---that 
are actually identified.  
Most of these are radio-loud sources, which represent only a minority
of AGNs \citep{Urr95}.
Broad-band X-ray surveys are not generally an effective method for finding
Type 2 quasars because even the X-ray continuum
may not  penetrate
the enshrouding gas. 
For example, CDF-S 202 is the
only high-luminosity obscured AGN reported
in the 1-Ms  exposure of the \chandra{} Deep Field South,
and only one has been identified in the 185-ks observation of the Lynx 
field \citep{Ste02}.

This sample suggests that  
prominent Fe K$\alpha$ (EW $> 1$ keV) is a common feature of 
very obscured AGNs, 
and the characteristic large EW can be exploited 
to find more of them. 
Specifically, we propose searching
for Type 2 quasars in continuum-subtracted 
{\em narrow-band} X-ray images.  
At high redshift, in particular,
where the host galaxy contribution is negligible,
the Fe line contains nearly all the X-ray flux.
Within a narrow energy range, $\Delta E \approx 300$ eV,
some objects might become significant detections that would
be undetectable when the background or sensitivity constraint
of a wider band is included.

The Fe K$\alpha$ line is a valuable probe of 
buried AGNs and the material that hides them.
Isolating this line emission both spectrally
and spatially from additional diluting sources,
these \chandra{} observations suggest that large EWs 
are common in obscured AGNs. 
The most extreme EWs require small torus opening angles
and arise in the starburst/AGN composite galaxies.

\acknowledgements
We thank Tahir Yaqoob for essential assistance with and software for
the grating spectra and thank David Strickland for providing a reduced
spectrum of the NGC 4945 nucleus.

\begin{deluxetable}{llllc}
\tabletypesize\scriptsize
\tablewidth{0pt}
\tablecaption{Spectral Model Parameters\label{tab:ew}}
\tablehead{
\colhead{Galaxy}
&\colhead{$E$\tablenotemark{a}}
&\colhead{EW\tablenotemark{a}}
&\colhead{$\Gamma$}
&\colhead{References}\\
&\colhead{(keV)}&\colhead{(keV)}
}
\startdata
M51       & $6.43\pm0.04$ & $4.9^{+2.5}_{-2.4}$    &$0.6^{+2.6}_{-0.3}$& 1, 2 \\
NGC 5135  & $6.39^{+0.03}_{-0.04}$ & $2.4^{+1.8}_{-0.5}$    &$0.0^{+1.6}_{-0.2}$& 3, 2 \\ 
Circinus  & $6.39\pm 0.07$         & $2.11\pm0.48$          &-0.8& 4, 5 \\   
NGC 4945  & $6.39\pm0.02$          & $1.26^{+0.26}_{-0.31}$ &$-0.98\pm0.7$ &6, 2 \\ 
NGC 1068  & $6.40$                 & $1.22$                 &$1.0^{+0.9}_{-1.1}$  &7, 8 \\ 
CDF-S 202 & $6.43\pm0.34$          & $1.19^{+1.2}_{-0.92}$  &1.8\tablenotemark{b}& 9 \\
IRAS 09104+4109 & $6.40\pm 0.07$   & $1.1^{+0.9}_{-0.5}$    &2.0\tablenotemark{b}& 10, 11\\
NGC 7130  & $6.40^{+0.05}_{-0.04}$ & $1.1^{+0.3}_{-0.5}$    &$-0.1^{+0.6}_{-0.2}$& 12, 2 \\ 
Mrk 3     & $6.39\pm0.01$          & $0.73^{+0.46}_{-0.37}$ &$1.9\pm0.8$ & 13, 2 \\ 
\enddata
\tablenotetext{a}{Source rest frame}
\tablenotetext{b}{Reflection model fixed intrinsic source}
\tablecomments{ 
Errors are 90\% confidence limits for one interesting parameter.}
\tablerefs{
(1) Makishima et al. (1990);
(2) this work;
(3) Turner et al. (1997); 
(4) Matt et al. (1996); 
(5) Sambruna et al. (2001);  
(6) Iwasawa et al. (1993);  
(7) Koyama et al. (1989); 
(8) Young et al. (2001); 
(9) Norman et al. (2002); 
(10) Fraceschini et al. (2000);
(11) Iwasawa et al. (2001); 
(12) Risaliti et al. (1999); 
(13) Awaki et al. (1990)
}
\end{deluxetable}

\clearpage


\end{document}